\documentclass[8pt,letterpaper]{article}
\usepackage[letterpaper, total={7.2in, 9.8in}]{geometry}
\usepackage{graphicx}
\usepackage{authblk}
\usepackage{hyperref}
\usepackage{amsmath} 
\usepackage{amssymb} 
\usepackage{orcidlink} 
\usepackage[numbers,super,sort,compress]{natbib}
\usepackage{balance} 
\usepackage{float}
\usepackage{xcolor}
\usepackage{url}
\hypersetup{
	colorlinks=true,
	linkcolor=blue,
	filecolor=magenta,
	urlcolor=blue,
	citecolor=blue,
	pdftitle={ECM-Signaling-WebApp},
	pdfauthor={Hasi Hays},
}
\makeatletter
\providecommand\doi[1]{\href{https://doi.org/#1}{\nolinkurl{#1}}}
\makeatother
\makeatletter
\renewcommand{\maketitle}{\bgroup\setlength{\parindent}{0pt}
	\begin{flushleft}
		\textbf{\@title}
		
		\@author
	\end{flushleft}\egroup}
\makeatother
\usepackage{caption}
\usepackage{titlesec}
\titleformat{\subsection}
{\normalsize\bfseries}
{\normalsize\thesubsection}
{1em}
{}
\usepackage[noabbrev]{cleveref}

\titleformat{\section}
{\normalsize\bfseries}
{\normalsize\thesection}
{1em}
{}
\let\oldsubsubsection\subsubsection
\renewcommand{\subsubsection}[1]{\oldsubsubsection{#1}}
\usepackage{booktabs}
\usepackage{changepage}
\usepackage{multicol}
\usepackage{fancyhdr}
\titlespacing*{\section}{0pt}{12pt}{6pt}

\title{
	\begin{center}
		\rule{\linewidth}{2pt}\\[0.5cm]
		{\Large \textbf{ECMSim: A high-performance interactive web application for real-time spatiotemporal simulation of cardiac ECM signaling and diffusion}}\\[0.3cm]
		\rule{\linewidth}{1pt}\\
		\vspace{0.7cm}
	\end{center}
}
\date{}
\author[1\orcidlink{0000-0003-0843-050X}] {\textbf{Hasi Hays}}
\author[1,*\orcidlink{0000-0001-8678-9716}] {\textbf{William J. Richardson}}
\affil[1]{\footnotesize Department of Chemical Engineering, University of Arkansas, Fayetteville, AR 72701, USA}
\affil[*]{\footnotesize Correspondence: \textcolor{blue}{wr013@uark.edu}}

\begin{document}
	\maketitle

\begin{adjustwidth}{0.5in}{0.5in}
\section*{\normalsize ABSTRACT}
Extracellular matrix (ECM) remodeling is central to a wide variety of healthy and diseased tissue processes. Unfortunately, predicting ECM remodeling under various chemical and mechanical conditions has proven to be excessively challenging, due in part to its complex regulation by intracellular and extracellular molecular reaction networks that are spatially and temporally dynamic. We introduce ECMSim, which is a highly interactive, real-time, and web application designed to simulate heterogeneous matrix remodeling. The current model simulates cardiac scar tissue with configurable input conditions using a large-scale model of the cardiac fibroblast signaling network. Cardiac fibrosis is a major component of many forms of heart failure. ECMSim solves 1.37 million coupled ordinary differential equations (ODEs) and executes approximately 4.84 million operations per time step in real time, encompassing 137 molecular species and 259 regulatory interactions per cell across a 100$\times$100 spatial array (10,000 cells), which accounts for inputs, receptors, intracellular signaling cascades, ECM production, feedback loops, and molecular diffusion. The algorithm is represented by a set of ODEs that are coupled with ECM molecular diffusion. The equations are solved on demand using compiled C++ and the WebAssembly standard. The platform includes brush-style cell selection to target a subset of cells with adjustable input molecule concentrations, parameter sliders to adjust parameters on demand, and multiple coupled real-time visualizations of network dynamics at multiple scales. Implementing ECMSim in standard web technologies enables a fully functional application that combines real-time simulation, visual interaction, and model editing. The software enables the investigation of pathological or experimental conditions, hypothetical scenarios, matrix remodeling, or the testing of the effects of an experimental drug(s) with a target receptor. 

\section*{\normalsize Keywords}

Matrix remodeling, Heart failure, Cardiac fibrosis, ODEs, Extracellular matrix, Spatial heterogeneity
\end{adjustwidth}

	\section*{\normalsize INTRODUCTION}

Extracellular matrix (ECM) remodeling represents one of the most computationally intensive and mechanistically complex biological processes in cardiovascular pathophysiology that require simultaneous integration of multi-scale reaction-diffusion dynamics, mechanochemical dynamics, and spatially heterogeneous cellular responses that span orders of magnitude in both temporal and spatial domains. \cite{Travers2016,Li2000} The inherent difficulty of simulating real-time matrix remodeling is due to the complex interdependencies between collagen synthesis and degradation kinetics, the regulation of matrix metalloproteinases (MMPs), and mechanical feedback loops that arise from the changing stiffness of the tissue and the orientation of the fiber. These interdependencies have been explored in various studies that investigate the roles of MMPs and their regulatory mechanisms in ECM homeostasis, particularly in pathological conditions such as myocardial infarction (MI) \cite{Buck2025} and idiopathic pulmonary fibrosis. \cite{Organ2019}  Moreover, the computational challenges posed by these dynamic interactions are well-documented, underscoring the need for advanced simulation techniques that can accommodate the intricacies of such biological systems. \cite{Zhang2016,Organ2019} The clinical importance of dysregulation of ECM is highlighted by the fact that MI remains one of the leading causes of death worldwide, accounting for an estimated 16 million deaths annually, based on data from the World Health Organization (WHO). \cite{Lu2022} Heart failure, which affects more than 64 million people worldwide, is characterized by significant matrix remodeling dynamics, driven in part by aging populations and improved survival rates from acute coronary events. \cite{Chen2025,Bertero2018,Kotta2023} The association between ECM alterations and disease progression reinforces the urgency of addressing these remodeling challenges. \cite{Edgar2018}

 Cardiac fibrosis plays an important role in heart failure, hypertrophic cardiomyopathy, and other heart illnesses. An excess of ECM proteins, especially collagens, is what makes this process happen. This excessive ECM accumulation changes the normal shape and function of the myocardium. \cite{Majid2023,Hall2021,Rogers2022} Cardiac fibroblasts are the primary cells that make the ECM, and they are responsive to multiple biophysical activities, like mechanical stress, inflammatory cytokines, growth factors, and neurohormonal signals. These signals work together in complicated ways to keep the balance between synthesis and degradation of the ECM. \cite{Rypdal2021,Sisto2021} The complexity of fibroblast signaling pathways presents key challenges for scientific studies, especially in matrix remodeling with multi-species diffusion. These pathways are interconnected and include transforming growth factor-$\beta$ (TGF-$\beta$) signaling, mitogen-activated protein kinase (MAPK) cascades, mechanotransduction pathways, and inflammatory mediator responses. \cite{Sisto2021,Kleinbongard2024} The dynamic interactions among these signaling pathways occur over various timescales, from immediate receptor activation to prolonged transcriptional changes that can last several hours to days. Furthermore, the spatial heterogeneity in cellular responses adds additional complexity, as does the intercellular communication facilitated by paracrine signaling. \cite{Larson2022} Traditional experimental approaches are limited in their ability to capture the full complexity of multi-pathway signaling dynamics combined with spatial molecular diffusion and feedback regulation. Computational modeling provides a complementary approach that can integrate these multiple interacting processes within a unified quantitative framework. 

In response of these shortcomings, computational modeling has become an important tool in cardiac fibrosis studies because it lets researchers combine different experimental results into complete models. These kinds of mathematical systems not only help come up with new hypotheses, but they also provide a way to test possible treatments. \cite{Wang2021,ImprotaCaria2024} Nonetheless, numerous current computational models of cardiac fibroblast signaling exhibit considerable deficiencies. A prevalent deficiency is that many of these models don't have the spatial resolution needed to accurately model how cells interact with each other and how tissues are organized. Most computational tools demand special software and knowledge, which makes it harder for more researchers to use them. \cite{Hall2021,AxelssonRaja2022} Recent improvements in web technologies, including WebAssembly, show promise for solving these problems. This technology lets web browsers do high-performance calculations at speeds that are similar to those of native apps, while also being easy to use and working on all platforms. \cite{Xu2022} These features are especially useful for scientific computing programs that need a lot of numerical analysis, which is very important in the investigation of cardiac fibrosis. This paper introduces ECMSim, a new web-based platform that simulates the signaling network of cardiac fibroblasts. This platform addresses the limitations of existing tools, including the lack of spatial resolution for cell-cell communication, the absence of real-time interactivity for parameter exploration, and the requirement for specialized software installation that limits accessibility, by using a comprehensive, system-level model that integrates all major signaling pathways implicated in cardiac fibroblast activation, from receptor-level inputs through intracellular signaling cascades to ECM protein outputs, within a single, unified computational framework in an easy-to-use interface. \cite{Hu2022} ECMSim consists of ordinary differential equations (ODEs) for mathematical representations of all the species organized into various functional modules that span receptor activation, inhibition, intracellular signaling cascades, transcriptional regulation, and intercellular feedback mechanisms. The combination of all the ODEs in all the cells are distributed in the grid with different altering input molecules and molecular diffusion in the spatial matrix. The computational engine harnesses optimized C++ code compiled to WebAssembly, delivering real-time performance for simulations involving large-scale spatial tissue grids containing 10,000 individual cells, each with a complete intracellular signaling network. \cite{Wang2021}

ECMSim's signaling network was informed by the logic-based ODE model of cardiac fibroblast signaling developed by Zeigler et al., \cite{Zeigler2016,Zeigler2016review} which demonstrated that a single-cell model could predict context-dependent drivers of myofibroblast differentiation with approximately 80\% accuracy across 82 input-output relationships. This mechanistic model has been well validated and extended in subsequent studies, including predictions of mechano-adaptive infarct therapies, \cite{Rogers2022b} computational screens for sex-specific drug effects in cardiac fibroblasts, \cite{Watts2023} and hierarchical molecular language models for signaling network analysis. \cite{Hays2025} Related models by Rogers et al.\ \cite{Rogers2022} and Rikard et al.\ \cite{Rikard2019} have explored fibroblast mechanotransduction and ECM regulation, respectively. However, these models are limited to single-cell representations without spatial coupling, lack real-time interactivity, and require specialized software for execution. ECMSim extends this foundation in several key ways: (1) expansion from a single-cell model to a spatially-resolved $100\times100$ tissue grid with reaction-diffusion coupling;  (2) implementation of real-time WebAssembly-based computation enabling interactive browser-based simulation; and (3) feedback loops enabling autocrine and paracrine signaling between cells.

By establishing a robust and user-friendly platform for real-time simulating cardiac fibroblast interactions with temporal and spatial dynamics, ECMSim significantly contributes to advancing the understanding of the molecular mechanisms underlying cardiac fibrosis and facilitates the exploration of novel therapeutic strategies. ECMSim is not limited to cardiac fibrotic matrix remodeling. It provides the foundation to model the advanced reaction-diffusion dynamics of specific applications, including cancer micro-environment, cell cross-talk, regenerative biology, wound healing, and drug discovery.

\captionsetup[figure]{labelformat=default}
\begin{figure}[!ht]
	\includegraphics[width=1\textwidth]{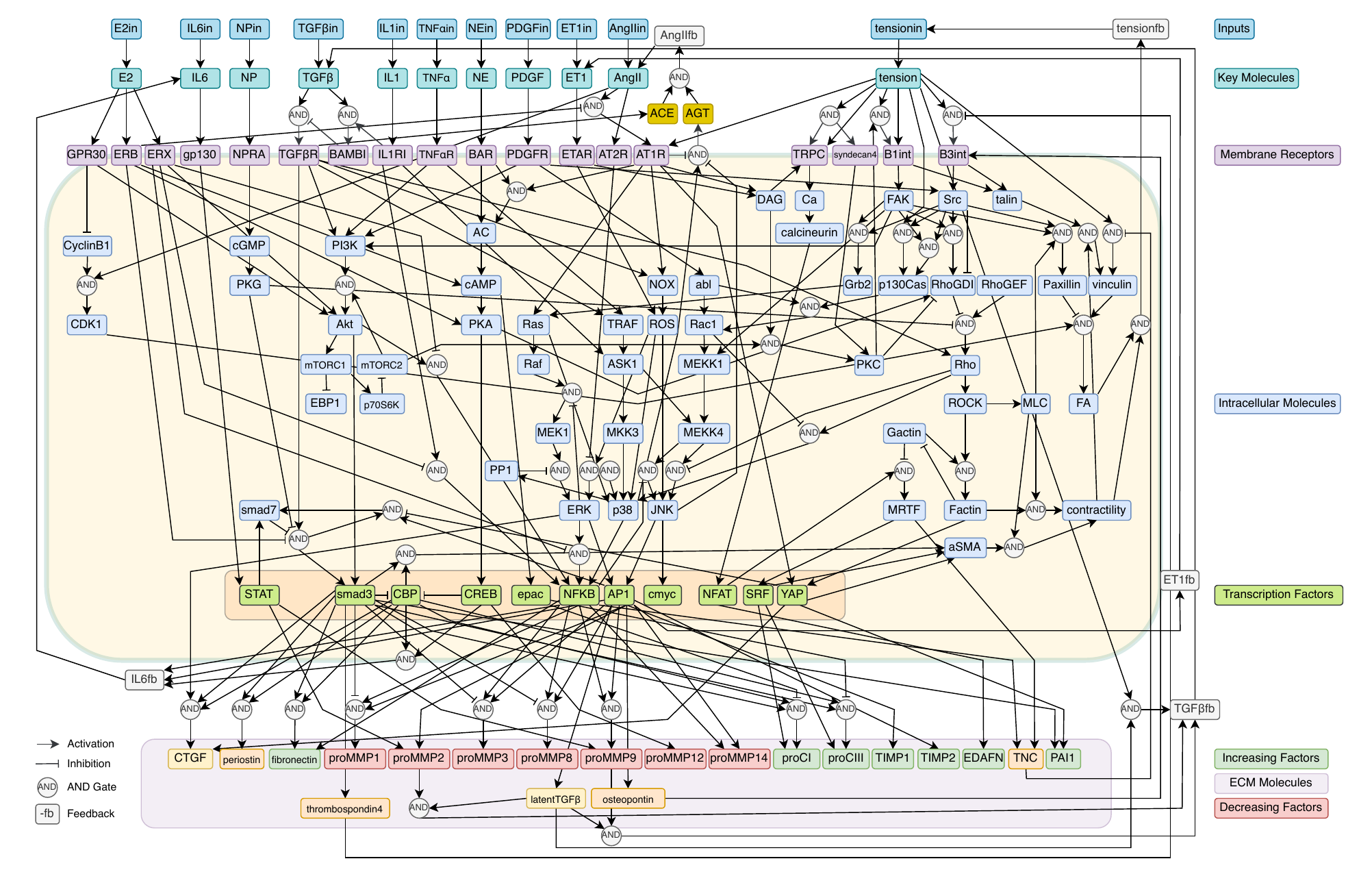}
	\caption{\footnotesize \textbf{Comprehensive cardiac fibroblast signaling network}}
	\footnotesize Network architecture of the cardiac fibroblast signaling model implemented in ECMSim. The network integrates ten input molecules (E2, IL6, NP, TGF-$\beta$, IL1, TNF-$\alpha$, NE, PDGF, ET1, AngII, and tension) that activate membrane receptors and trigger intracellular signaling cascades. Key signaling modules include MAPK pathways (ERK, p38, JNK), PI3K-Akt-mTORC signaling, Rho GTPase mechanotransduction, and calcium-dependent pathways. These converge on transcription factors (STAT, Smad3, CREB, AP1, NFAT, SRF, YAP, NF-$\kappa$B) that regulate expression of ECM molecules including collagens (proCI, proCIII), matrix metalloproteinases (proMMP1, 2, 3, 8, 9, 12, 14), tissue inhibitors (TIMP1, TIMP2), matricellular proteins (CTGF, periostin, fibronectin, tenascin-C, thrombospondin-4, osteopontin), and other ECM regulators ($\alpha$-SMA, PAI1, EDAFN). Feedback molecules (TGF-$\beta_{\text{fb}}$, AngII$_{\text{fb}}$, IL6$_{\text{fb}}$, ET1$_{\text{fb}}$, tension$_{\text{fb}}$) enable autocrine and paracrine signaling. Arrows indicate activation, bar-headed lines indicate inhibition, and ``AND'' gates represent cooperative regulatory interactions requiring multiple inputs for activation.
	\label{Figure 1}
\end{figure}
	
\section*{\normalsize METHODOLOGY}

\subsection*{\normalsize Mathematical model and system overview}

The ECMSim platform models cardiac fibroblast signaling through a comprehensive system of ordinary differential equations (ODEs) describing the temporal evolution of molecular species concentrations across a spatially-resolved tissue grid. The complete system encompasses 132 molecular species per cell (115 intracellular signaling species and 17 extracellular ECM species) distributed across a $100 \times 100$ cellular grid, yielding $1.32 \times 10^6$ intracellular variables. Including five diffusible feedback signals (TGF-$\beta_{\text{fb}}$, AngII$_{\text{fb}}$, IL6$_{\text{fb}}$, ET1$_{\text{fb}}$, and tension$_{\text{fb}}$) across the spatial domain adds $5 \times 10^4$ additional variables, resulting in $1.37 \times 10^6$ coupled differential equations solved simultaneously at each time step. The signaling network contains 259 regulatory interactions (edges), of which approximately 216 are activating and 43 are inhibitory. Table~\ref{Table 3} summarizes the distribution of molecular species across functional modules.

\begin{table}[!ht]
\centering
\footnotesize
\caption{\footnotesize \textbf{Distribution of molecular species across functional modules in ECMSim}}
\begin{tabular}{lrl}
\toprule
\textbf{Module} & \textbf{Species} & \textbf{Examples} \\
\midrule
Ligands/Inputs & 11 & AngII, TGF-$\beta$, IL6, ET1, tension \\
Receptors & 10 & AT1R, TGFB1R, ETAR, PDGFR, gp130 \\
Second messengers & 7 & ROS, DAG, cAMP, cGMP, Ca$^{2+}$ \\
Kinases/Phosphatases & 5 & PKA, PKC, PKG, calcineurin, PP1 \\
MAPK pathway & 11 & Ras, Raf, MEK1, ERK, p38, JNK \\
PI3K-Akt-mTOR & 6 & PI3K, Akt, mTORC1, mTORC2, p70S6K \\
Rho/ROCK \& cytoskeleton & 20 & Rho, ROCK, FAK, Src, YAP, Factin \\
Transcription factors & 8 & NF-$\kappa$B, AP1, STAT, Smad3, SRF, NFAT \\
Estrogen signaling & 5 & ER$\alpha$, ER$\beta$, GPR30, CyclinB1 \\
Additional regulatory & 14 & AGT, ACE, BAMBI, $\alpha$-SMA, LOX \\
ECM precursors (intracellular) & 17 & proCI, proCIII, proMMP2, TIMP1 \\
\midrule
\textit{Intracellular subtotal} & \textit{115} & \\
ECM molecules (extracellular) & 17 & proCI, fibronectin, proMMP9, TIMP2 \\
Feedback molecules & 5 & TGF-$\beta_{\text{fb}}$, AngII$_{\text{fb}}$, tension$_{\text{fb}}$ \\
\midrule
\textbf{Total per cell} & \textbf{137} & \\
\textbf{Total on grid} ($100\times100$) & \textbf{$1.37 \times 10^6$} & \\
\bottomrule
\end{tabular}
\label{Table 3}
\end{table}

Each molecular species $X_i$ is represented as a continuous variable whose dynamics are governed by production, degradation, and regulatory interaction processes. The general mathematical framework follows mass action kinetics with multiplicative regulatory interactions:

\begin{equation}
	\frac{dX_i}{dt} = \sum_{j} k_{j}^{\text{prod}} \prod_{k \in \text{activators}} X_k - \sum_{l} k_{l}^{\text{inhib}} X_i \prod_{m \in \text{inhibitors}} X_m - k_i^{\text{deg}} X_i
\end{equation}
where the first term represents production processes activated by upstream signaling molecules, the second term represents inhibition processes where molecule $X_i$ is suppressed by inhibitory factors, and the third term represents first-order degradation. All molecular concentrations are constrained to physiologically meaningful bounds through numerical clamping:

\begin{equation}
	X_i(t + \Delta t) = \max(0, \min(1, X_i(t + \Delta t)))
\end{equation}
ensuring concentrations remain within the normalized range $[0,1]$, where 0 represents an inactive state and 1 represents maximal physiological activity. This clamping serves as a numerical safeguard against transient overshoots or negative values that the forward Euler integration can produce during a single time step, particularly when multiple strong activation signals converge simultaneously, while also providing a consistent and intuitive framework for comparing activity levels across diverse molecular species with vastly different absolute concentrations.

\captionsetup[figure]{labelformat=default}
\begin{figure}[!ht]
	\includegraphics[width=1\textwidth]{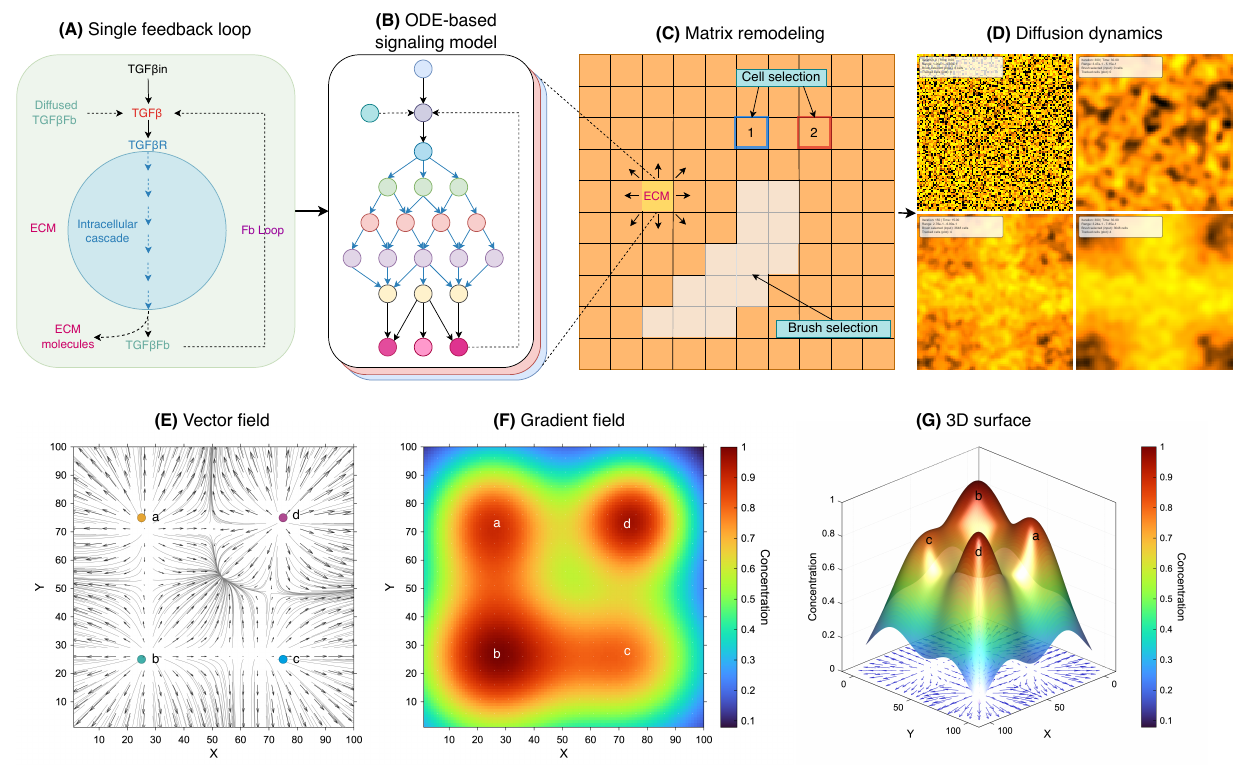}
	\caption{\footnotesize \textbf{ECMSim network architecture and diffusion model.}}
	\footnotesize (\textbf{A}) Single feedback loop: TGF-$\beta$ input activates its receptor (TGF-$\beta$R), which triggers intracellular signaling cascades that produce both ECM molecules and the feedback molecule TGF-$\beta_{\text{fb}}$; the feedback molecule diffuses to neighboring cells, completing the autocrine/paracrine loop. (\textbf{B}) ODE-based signaling model: hierarchical network architecture showing crosstalk between signaling modules including MAPK cascades (ERK, p38, JNK), PI3K-Akt-mTOR, Rho GTPase, and mechanotransduction pathways, from input receptors (top) through intracellular kinases to ECM outputs (bottom). (\textbf{C}) Matrix remodeling: the cellular grid representation with cell selection (cells 1 and 2 for temporal tracking), brush-based spatial selection of cell subpopulations, and bidirectional ECM molecule exchange between neighboring cells. (\textbf{D}) Diffusion dynamics: four representative simulation snapshots showing the temporal evolution of molecular concentrations from random initialization (top-left, iteration 0) through intermediate diffusion (top-right, iteration 300 without brush-selected input) to the equilibrated state without stimulation (bottom-left, iteration 150) and with localized stimulation (bottom-right, iteration 300), demonstrating how brush-applied inputs drive spatially heterogeneous concentration patterns. (\textbf{E}) Vector field streamlines illustrating the direction and magnitude of diffusive flux for four discrete concentration sources (points a, b, c, d) placed at different grid locations. (\textbf{F}) Gradient field showing the resulting two-dimensional concentration distribution from the four sources, with the color scale indicating normalized concentration (0--1). (\textbf{G}) Three-dimensional surface visualization of the same concentration field, where peak height corresponds to local concentration at each source location, illustrating how multiple diffusion sources interact to form the composite spatial concentration landscape.
	\label{Figure 2}
\end{figure}

\subsection*{Intracellular signaling network architecture}

The signaling network (Fig.~\ref{Figure 1}) architecture organizes molecular species into functionally distinct yet interconnected modules (receptors, second messengers, kinases, transcription factors, and ECM outputs) that reflect the hierarchical organization of biological signaling pathways, from extracellular inputs through intracellular processing to effector outputs, while facilitating extensive crosstalk among pathways. The input processing module handles ten primary signaling molecules including TGF-$\beta$, angiotensin II (AngII), interleukin-6 (IL6), interleukin-1 (IL1), tumor necrosis factor-$\alpha$ (TNF-$\alpha$), norepinephrine (NE), platelet-derived growth factor (PDGF), endothelin-1 (ET1), natriuretic peptides (NP), and estrogen (E2). Each input signal incorporates both external stimuli and intercellular feedback according to:

\begin{equation}
	\frac{dX_{\text{signal}}}{dt} = k_{\text{input}} I_{\text{external}} + k_{\text{feedback}} C_{\text{feedback}} - k_{\text{deg}} X_{\text{signal}}
\end{equation}
where $I_{\text{external}}$ represents user-defined input concentrations applied to selected cell populations through the brush selection interface, $C_{\text{feedback}}$ represents diffusible signals from neighboring cells, and rate constants $k_{\text{input}}$, $k_{\text{feedback}}$, and $k_{\text{deg}}$ govern input processing, feedback sensitivity, and degradation, respectively (Fig.~\ref{Figure 2}A).

The receptor activation module models the dynamics of ten major receptor classes including G-protein coupled receptors, receptor tyrosine kinases, cytokine receptors, and mechanosensitive ion channels. Receptor activation incorporates competitive inhibition through multiplicative regulatory terms:

\begin{equation}
	\frac{dR_{\text{active}}}{dt} = k_{\text{receptor}} L \cdot R_{\text{total}} - k_{\text{inhib}} R_{\text{active}} \cdot I_{\text{inhibitor}} - k_{\text{deg}} R_{\text{active}}
\end{equation}
where $L$ represents ligand concentration, $R_{\text{total}}$ is the total receptor pool, $I_{\text{inhibitor}}$ represents endogenous inhibitory molecules such as BAMBI for TGF-$\beta$ signaling or estrogen receptor $\beta$ for angiotensin signaling, and rate constants control activation, inhibition, and degradation processes.

Second messenger systems transduce receptor activation into intracellular signals through eight key mediators, including cyclic nucleotides (cAMP, cGMP), lipid mediators (diacylglycerol), calcium signaling, and reactive oxygen species. The protein kinase and phosphatase networks integrate second messenger signals through fifteen kinases and phosphatases, including protein kinase A, protein kinase C, and various mitogen-activated protein kinases. Downstream signaling cascades exhibit complex regulatory patterns through multiplicative interactions:

\begin{equation}
	\frac{dK_{\text{active}}}{dt} = k_{\text{act}} U_{\text{kinase}} \cdot S_{\text{substrate}} - k_{\text{phos}} K_{\text{active}} \cdot P_{\text{phosphatase}} - k_{\text{deg}} K_{\text{active}}
\end{equation}
where $U_{\text{kinase}}$ represents upstream kinase activity, $S_{\text{substrate}}$ is the inactive substrate pool, $P_{\text{phosphatase}}$ represents phosphatase activity, and rate constants control activation, dephosphorylation, and degradation rates.

The MAPK signaling module includes detailed representations of the ERK1/2, p38, and JNK pathways with twelve pathway components and their characteristic regulatory mechanisms. The transcriptional regulation module represents eight major transcription factor families, including nuclear factor $\kappa$B, activator protein 1, signal transducer and activator of transcription proteins, serum response factor, nuclear factor of activated T-cells, cAMP response element-binding protein, and Smad proteins. Mechanotransduction pathways convert mechanical forces into biochemical signals through eighteen cytoskeletal components, including integrin-mediated focal adhesion complexes, Rho family GTPase signaling, and cytoskeletal remodeling elements. The extracellular matrix production module represents seventeen key matrix proteins, including procollagen types I and III, fibronectin, periostin, tenascin-C, matrix metalloproteinases, and tissue inhibitors of metalloproteinases.

\captionsetup[figure]{labelformat=default}
\begin{figure}[!ht]
	\includegraphics[width=1\textwidth]{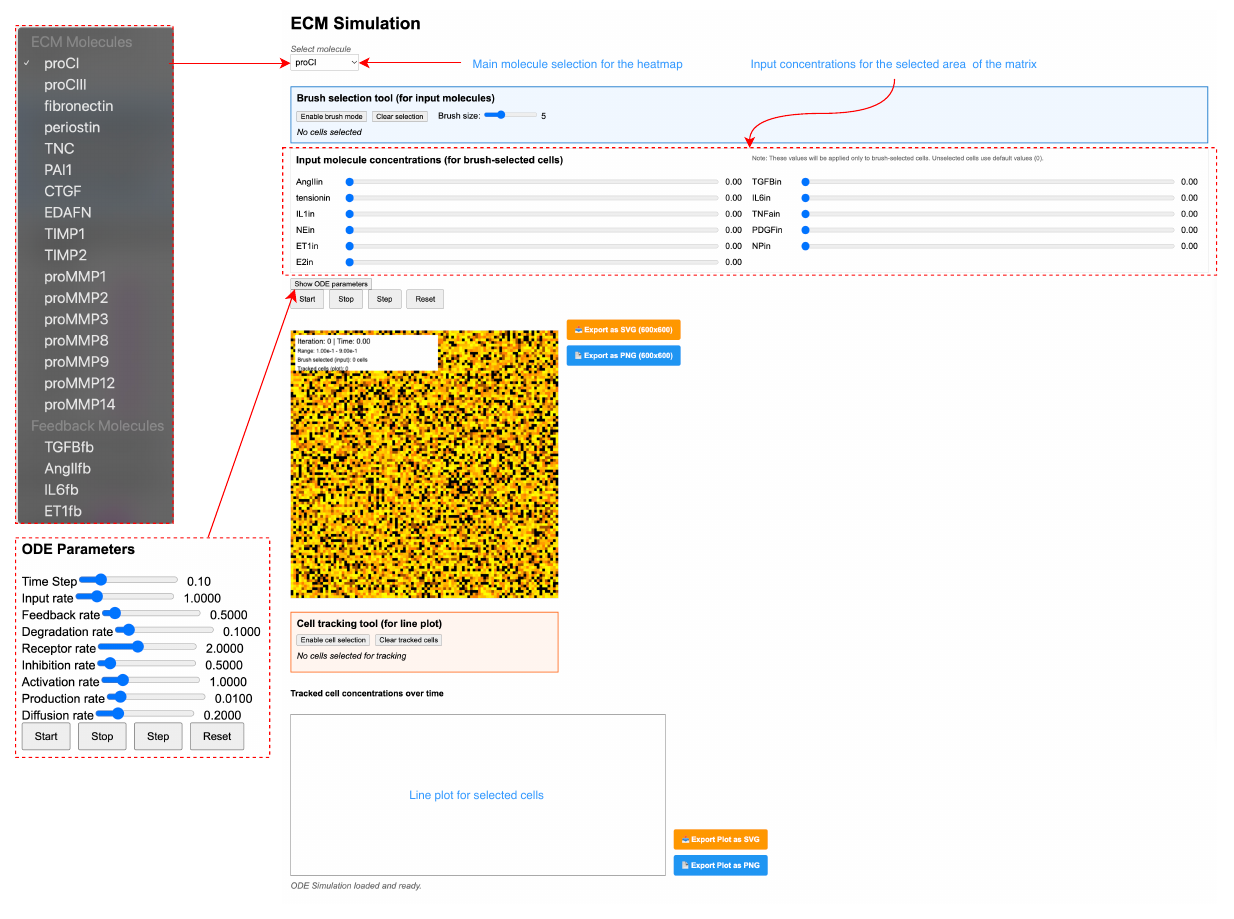}
	\caption{\footnotesize \textbf{Overview of the ECMSim web application.}}
	\footnotesize The interface comprises several interconnected panels. \textit{Left:} Molecule selection sidebar listing all 17 ECM molecules and 4 feedback molecules available for spatial visualization. \textit{Top:} The ``Select molecule'' dropdown determines which molecular species is displayed on the main heatmap; the input molecule concentration sliders (AngII, TGF-$\beta$, IL6, tension, IL1, TNF-$\alpha$, NE, PDGF, ET1, NP, E2) set stimulus levels applied exclusively to brush-selected cells. \textit{Center:} The brush selection tool enables spatial targeting of cell subpopulations with adjustable brush size (1--15); simulation controls (Start, Stop, Step, Reset) govern execution; the main $100\times100$ heatmap displays real-time concentration dynamics with iteration count, simulation time, and concentration range. Export buttons allow saving the heatmap as SVG or PNG. \textit{Bottom:} The cell tracking tool enables selection of up to 8 individual cells for temporal monitoring, with their concentration trajectories plotted in real time; tracked cell plots can also be exported. \textit{Bottom-left:} The ODE parameters panel provides real-time adjustment of all nine primary rate constants (time step, input, feedback, degradation, receptor, inhibition, activation, production, and diffusion rates) via interactive sliders.
	\label{Figure 3}
\end{figure}

\subsection*{\normalsize Spatial modeling and cellular grid architecture}

ECMSim implements spatial organization (Fig.~\ref{Figure 2}) through a regular two-dimensional grid of $100 \times 100$ individual cells, where each grid site represents a single cardiac fibroblast. The grid spacing $\Delta x \approx 10\,\mu$m corresponds to the typical diameter of a cardiac fibroblast (10--20\,$\mu$m), so the full grid represents a tissue region of approximately 1\,mm $\times$ 1\,mm. Each grid location $(i,j)$ contains a complete representation of the intracellular signaling network, creating a spatially-resolved model with over 1.3 million individual molecular species variables distributed across the computational domain. For larger tissue regions, the grid dimensions can be increased (e.g., $200\times200$ or $500\times500$), with computational cost scaling linearly with the number of grid cells.

Spatial coupling between cells is implemented through the diffusion of five key \textit{feedback signaling molecules} that mediate paracrine signaling and coordinate tissue-level responses. These are small, rapidly diffusing molecules, each with its own diffusion coefficient $D_k$, reflecting their distinct molecular sizes and tissue transport properties. The spatial evolution follows a discrete diffusion equation using finite difference approximation on the cellular grid:

\begin{equation}
	\frac{dC_{i,j}^{(k)}}{dt} = D_k \sum_{(m,n) \in \mathcal{N}_{i,j}} \left(C_{m,n}^{(k)} - C_{i,j}^{(k)}\right) + P_{i,j}^{(k)} - \lambda_k C_{i,j}^{(k)}
\label{Eq6}
\end{equation}
where $C_{i,j}^{(k)}$ represents the concentration of feedback molecule $k$ at grid position $(i,j)$, $\mathcal{N}_{i,j}$ denotes the eight-connected neighborhood of cell $(i,j)$, $D_k$ is the molecule-specific diffusion coefficient, $P_{i,j}^{(k)}$ represents the cellular production rate that depends on the intracellular signaling state of the cell at position $(i,j)$ (i.e., $P_{i,j}^{(k)}$ is a function of relevant intracellular precursor concentrations), and $\lambda_k$ represents extracellular degradation or clearance.

The neighborhood operator for the discrete Laplacian includes all eight adjacent cells:

\begin{equation}
	\mathcal{N}_{i,j} = \{(i \pm 1, j \pm 1), (i \pm 1, j), (i, j \pm 1)\} \bmod N
\end{equation}
where $N = 100$ is the grid size and the modulo operation implements periodic boundary conditions such that cells at grid boundaries interact with cells on the opposite edge to minimize finite-size effects.

The five feedback molecules include secreted TGF-$\beta$ that acts in autocrine and paracrine modes to amplify fibrotic responses, angiotensin II produced through local renin-angiotensin system activation, IL6 that coordinates inflammatory responses, endothelin-1 that propagates vasoconstrictor and pro-fibrotic signals, and biomechanical signal (tension$_{\text{fb}}$)  an indirect way to capture force transmission that can mechanically stimulate neighboring cells across a distance. In contrast to the feedback signaling molecules described above, \textit{ECM and ECM-binding molecules} (collagens, fibronectin, periostin, etc.) are large, locally deposited proteins. Their concentrations evolve according to production from activated fibroblasts and limited spatial diffusion reflecting their larger molecular size and tendency for local deposition:

\begin{equation}
	\frac{dE_{i,j}^{(m)}}{dt} = k_{\text{prod}} X_{i,j}^{\text{precursor}(m)} - k_{\text{deg}} E_{i,j}^{(m)} + 0.2 \cdot D_{\text{feedback}} \sum_{(p,q) \in \mathcal{N}_{i,j}} \left(E_{p,q}^{(m)} - E_{i,j}^{(m)}\right)
\end{equation}
where $E_{i,j}^{(m)}$ represents ECM molecule $m$ at position $(i,j)$, $X_{i,j}^{\text{precursor}(m)}$ is the corresponding intracellular precursor concentration, and ECM molecules diffuse at 20\% the rate of feedback molecules to reflect their distinct biophysical properties. The 0.2 scaling factor is grounded in biophysical evidence: ECM structural proteins such as collagens (100--400\,kDa) and fibronectin ($\sim$440\,kDa) are substantially larger than feedback signaling molecules such as TGF-$\beta$ ($\sim$25\,kDa), IL6 ($\sim$21\,kDa), and AngII ($\sim$1\,kDa). According to Milo and Phillips, \cite{Milo2015} diffusion coefficients decrease by roughly one order of magnitude per order-of-magnitude increase in molecular weight, yielding a ratio of approximately 0.1--0.2 for large versus small proteins. This size-dependent hindrance has been directly measured in collagen gels using fluorescence recovery after photobleaching (FRAP) \cite{Ramanujan2002} and modeled through steric, hydrodynamic, and electrostatic interactions in the ECM. \cite{Stylianopoulos2010}

\subsection*{Numerical integration and computational methods}

The complete system of $1.37 \times 10^6$ coupled ODEs is integrated using the forward Euler method with adaptive time-stepping to maintain numerical stability while maximizing computational efficiency:

\begin{equation}
	X_i(t + \Delta t) = X_i(t) + \Delta t \cdot \frac{dX_i}{dt}\bigg|_{t}
\end{equation}
The time step $\Delta t$ is dynamically adjusted based on the maximum rate of change across all molecular species to ensure numerical stability. While higher-order integration methods such as Runge-Kutta algorithms offer improved accuracy, the forward Euler approach provides several advantages for interactive simulation applications, including computational efficiency and numerical stability for the specific rate constants employed, with predictable execution times that facilitate real-time visualization updates. All molecular concentrations are subject to physiological constraints at each integration step, ensuring values remain within normalized bounds, representing meaningful biological activity levels. The spatial derivatives in the diffusion equations are approximated using second-order finite difference methods with periodic boundary conditions to minimize edge effects during typical simulation scenarios.

\subsection*{\normalsize Parameter estimation and model calibration}

The model incorporates eight primary rate constants that can be adjusted (from the default set value) in real-time through the user interface, as summarized in Table~\ref{Table 2}. As the model uses normalized concentrations in the range $[0,1]$ representing fractional activation levels, all rate constants have units of inverse time (time step$^{-1}$). In the discrete diffusion formulation (\autoref{Eq6}), the grid spacing $\Delta x$ is already absorbed into the diffusion coefficient through the finite difference discretization, so $D_k = D_{\text{physical}} / \Delta x^2$.

\begin{table}[!ht]
\centering
\footnotesize
\caption{\footnotesize \textbf{Primary adjustable rate constants in ECMSim}}
\begin{tabular}{llll}
\toprule
\textbf{Parameter} & \textbf{Description} & \textbf{Default} & \textbf{Units} \\
\midrule
$k_{\text{input}}$ & Input processing rate & 1.0 & time step$^{-1}$ \\
$k_{\text{feedback}}$ & Feedback sensitivity & 0.5 & time step$^{-1}$ \\
$k_{\text{degradation}}$ & Degradation rate & 0.1 & time step$^{-1}$ \\
$k_{\text{receptor}}$ & Receptor activation rate & 2.0 & time step$^{-1}$ \\
$k_{\text{inhibition}}$ & Inhibition rate & 0.5 & time step$^{-1}$ \\
$k_{\text{activation}}$ & Activation rate & 1.0 & time step$^{-1}$ \\
$k_{\text{production}}$ & ECM production rate & 0.01 & time step$^{-1}$ \\
$k_{\text{diffusion}}$ & Diffusion rate & 0.25 & time step$^{-1}$ \\
\bottomrule
\end{tabular}
\label{Table 2}
\end{table}

\subsection*{\normalsize Computational implementation and optimization}

The computational core of ECMSim is implemented in C++ using object-oriented design principles with optimized data structures, including unordered maps for efficient molecular species lookup and vector arrays for spatial grid representation. Memory allocation is managed through pre-allocation strategies that minimize dynamic memory operations during simulation execution. The simulation state is organized to maximize cache locality by storing related molecular species in contiguous memory locations. The C++ simulation code is compiled to WebAssembly using the Emscripten toolchain with aggressive optimization flags including function inlining, loop unrolling, and dead code elimination. The resulting WebAssembly binary achieves computational performance within approximately 20\% of equivalent native C++ code while maintaining complete cross-platform compatibility across modern web browsers. The WebAssembly linear memory model provides direct access to simulation data from JavaScript visualization code, eliminating costly data serialization and transfer operations. Memory management within the WebAssembly environment utilizes several optimization strategies: simulation state organization for cache locality, temporary variable reuse across integration steps, and leveraging the linear memory model for efficient data access. The platform executes efficiently on standard computing hardware, requiring only a modern web browser with WebAssembly support for typical simulations involving the full 1.37 million variable system.

\subsection*{\normalsize Performance benchmarks}

Performance profiling of ECMSim was conducted using Chrome DevTools on a MacBook Pro (14-inch, 2021; Apple M1 Max; 32\,GB RAM; Google Chrome). Table~\ref{Table 1} summarizes the key performance metrics measured during real-time simulation of the full $100 \times 100$ grid with 1.37 million coupled ODEs.

\begin{table}[H]
\centering
\footnotesize
\caption{\footnotesize \textbf{Performance benchmarks for ECMSim during real-time simulation}}
\begin{tabular}{p{6.5cm}r}
\toprule
\textbf{Metric} & \textbf{Measured Value} \\
\midrule
\multicolumn{2}{l}{\textbf{Computational scale}} \\
Total coupled ODEs & $1.37 \times 10^6$ \\
Spatial grid & $100 \times 100$ (10,000 cells) \\
Species per cell & 137 \\
Regulatory interactions (edges) per cell & 259 \\
\midrule
\multicolumn{2}{l}{\textbf{Operations per time step}} \\
\hspace{0.3cm}ODE rate evaluations ($137 \times 10{,}000$) & 1,370,000 \\
\hspace{0.3cm}Euler integration updates & 1,490,000 \\
\hspace{0.3cm}Feedback diffusion ($5 \times 10{,}000 \times 8$ neighbors) & 450,000 \\
\hspace{0.3cm}ECM diffusion ($17 \times 10{,}000 \times 8$ neighbors) & 1,530,000 \\
\textbf{Total operations per time step} & $\mathbf{\sim 4.84 \times 10^6}$ \\
\midrule
\multicolumn{2}{l}{\textbf{Memory and runtime}} \\
Total memory footprint (browser tab) & $\sim$523\,MB \\
JS Heap memory & 37.2--42.2\,MB \\
CPU time: Scripting (ODE solver) & 99.4\% \\
CPU time: Painting (canvas rendering) & 0.25\% \\
CPU time: Rendering (layout) & 0.17\% \\
CPU time: System overhead & 0.15\% \\
CPU utilization & $\sim$105\% (single-core) \\
Minimum recommended RAM & 4\,GB \\
Simulation mode & Real-time interactive \\
Browser compatibility & Chrome, Firefox, Safari \\
\bottomrule
\end{tabular}
\label{Table 1}
\end{table}

Each simulation time step executes approximately 4.84 million operations, comprising 1.37 million ODE rate evaluations, 1.49 million Euler integration updates, and 1.98 million diffusion neighbor computations across the feedback (5 molecules) and ECM (17 molecules) spatial grids. The total browser memory footprint of approximately 523\,MB (including WebAssembly linear memory, JavaScript heap, and canvas rendering buffers) is modest for a system of this scale, and the JavaScript heap alone requires only 37--42\,MB. The CPU profiling confirms that over 99\% of computation time is dedicated to the ODE integration, with negligible overhead for visualization rendering ($<$0.5\%), indicating efficient allocation of computational resources to the numerical simulation core. The platform has been tested across Chrome, Firefox, and Safari, with Chrome providing the best WebAssembly performance due to its optimized V8 engine.

\captionsetup[figure]{labelformat=default}
\begin{figure}[!ht]
	\includegraphics[width=1\textwidth]{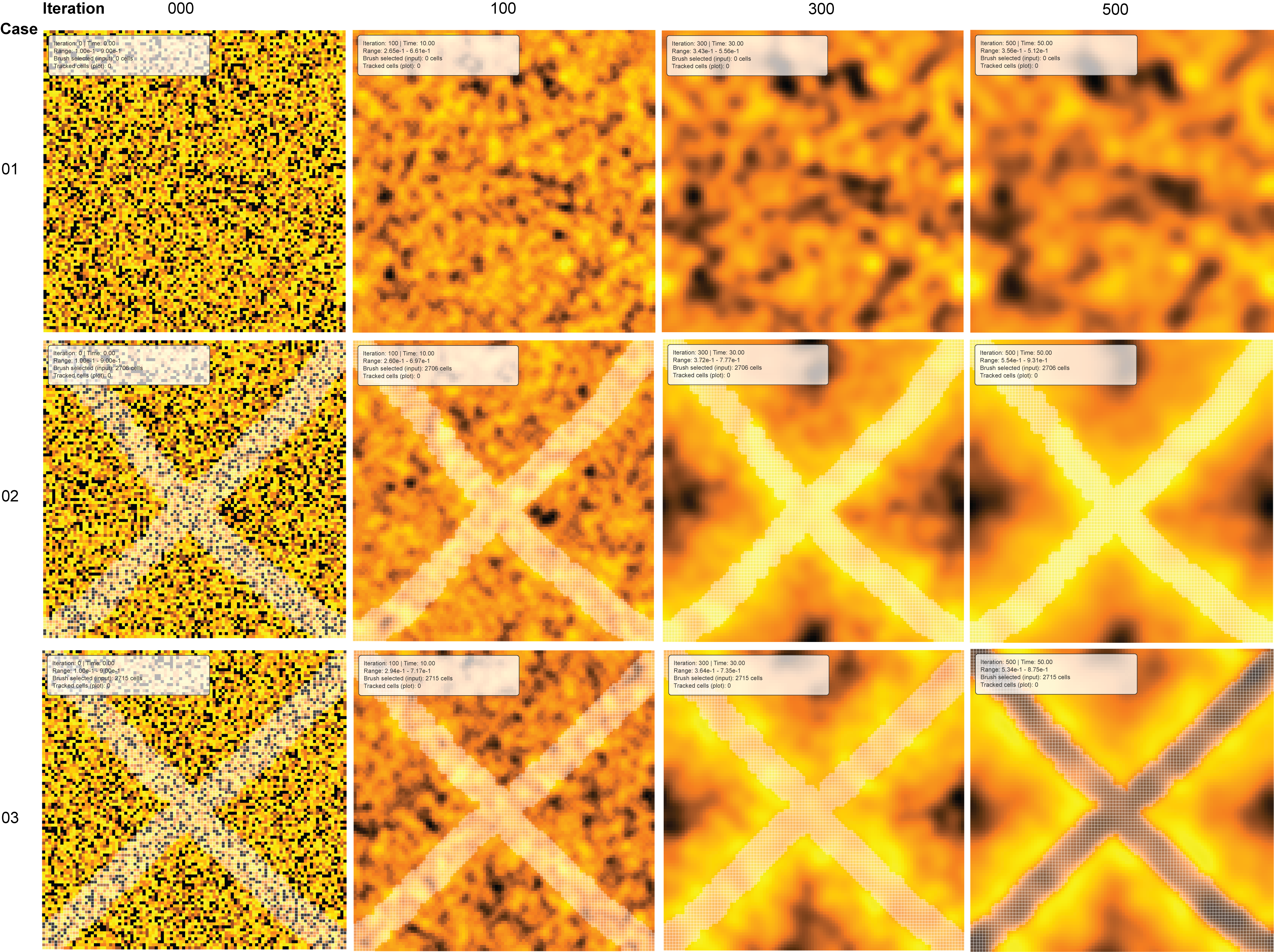}
	\caption{\footnotesize \textbf{Diffusion dynamics}}
	\footnotesize Screenshots of 3 distinct cases of simulation. All the matrices have collagen I in the ECM and simulated 4 different iterations (0, 100, 300, 500) with 0.1 time steps (S0, simulation times 0, 10, 30, and 50 respectively). More yellow color represents the higher concentration and black color represents the lowest concentration in the matrix. Case \textbf{01}: No brush selected. So, the inputs molecules have default values (all at 0). Case \textbf{02}: Brush selected (X shape). TGF-$\beta$ input value is 1 in the brushed area. Rest of the inputs are in default values (0).  Unselected area has default input molecule concentration. Case \textbf{03}: Brush selected (X shape). TGF-$\beta$, AngII, IL6, IL1, TNF$\alpha$, and NE input molecules are set at 1. Unselected area has default input molecule concentration. Each brush selected cases (2 and 3 - each case separately) included same brush selection throughout the continuous temporal iterations.
	\label{Figure 4}
\end{figure}
	
\section*{\normalsize RESULTS}

\subsection*{\normalsize Integrated signaling network architecture and spatial dynamics}

ECMSim successfully implements a comprehensive computational framework for simulating cardiac ECM remodeling through the integration of 137 molecular species and 259 regulatory interactions distributed across a spatially-resolved tissue grid. Fig.~\ref{Figure 2}A illustrates the fundamental feedback loop architecture underlying the model, where external TGF-$\beta$ input triggers intracellular signaling cascades that ultimately produce both ECM molecules and feedback molecules that diffuse to neighboring cells. This autocrine and the paracrine signaling mechanism enables the propagation and amplification of fibrotic responses through the tissue. The complete signaling network architecture (Fig.~\ref{Figure 2}B) encompasses the major pathways implicated in cardiac fibrosis, including TGF-$\beta$/Smad signaling, MAPK cascades, PI3K-Akt-mTOR pathways, and mechanotransduction networks. The hierarchical organization of the network enables signal integration from multiple upstream inputs through intermediate processing nodes to downstream effector molecules responsible for matrix protein synthesis. The modular architecture facilitates systematic investigation of pathway crosstalk and the identification of critical regulatory nodes that control fibrotic responses.

The spatial implementation (Fig.~\ref{Figure 2}C) demonstrates the platform's capability to model tissue-level organization through a regular cellular grid where each cell contains the complete intracellular signaling network. Intercellular communication is mediated through the diffusion of five key feedback molecules that coordinate regional responses and enable the emergence of spatially heterogeneous patterns. The diffusion dynamics (Fig.~\ref{Figure 2}D) show representative simulation outputs displaying the spatial distribution of molecular concentrations at successive time points, illustrating how localized stimuli can propagate across the tissue domain through diffusive coupling. To further characterize the spatial diffusion behavior, Fig.~\ref{Figure 2}E--G presents vector field streamlines, gradient fields, and three-dimensional surface visualizations for four discrete concentration sources placed at different grid locations. The vector field (Fig.~\ref{Figure 2}E) illustrates the direction and magnitude of diffusive flux, while the gradient field (Fig.~\ref{Figure 2}F) and 3D surface (Fig.~\ref{Figure 2}G) show how multiple diffusion sources interact to form composite concentration landscapes with smooth spatial gradients.

In ECMSim, the fibrotic response can be quantified through multiple complementary metrics: (1) collagen accumulation, specifically the concentrations of procollagen type~I (proCI) and procollagen type~III (proCIII), which are the primary structural ECM proteins in cardiac fibrosis; (2) the collagen/MMP balance, representing the ratio of collagen production to MMP-mediated degradation; (3) myofibroblast activation markers, particularly $\alpha$-SMA expression; and (4) overall ECM protein accumulation across all 17 modeled species. While collagen accumulation is the most commonly used single metric for fibrosis severity, the platform enables multi-dimensional characterization of the fibrotic response, allowing researchers to select the most appropriate output variables for their specific research questions.

\subsection*{\normalsize Interactive simulation platform and real-time parameter control}

The ECMSim web application provides an intuitive interface for exploring complex signaling dynamics through multiple integrated visualization modalities (Fig.~\ref{Figure 3}). The primary heatmap visualization displays the spatial distribution of selected molecular species across the 100×100 cellular grid, with real-time updates during simulation execution that enable immediate observation of evolving concentration patterns. The molecule selection panel allows users to dynamically switch between 17 ECM molecules and 4 feedback molecules, instantly updating both spatial visualizations and temporal plots without requiring simulation restart. The innovative brush selection tool enables precise spatial control over input conditions by allowing users to define arbitrary cell populations for targeted interventions. The brush interface supports adjustable sizing from single-cell precision to broad regional selections, accommodating diverse experimental scenarios ranging from localized growth factor application to regional mechanical stimulation. The input concentration control panel provides independent adjustment of 10 primary signaling molecules, with real-time value displays and immediate propagation to selected cell populations. This design enables complex experimental scenarios such as spatially heterogeneous inflammation conditions or combinatorial cytokine treatments.

The visualization system combines tissue-level spatial patterns with detailed single-cell temporal dynamics. Users can select up to eight individual cells directly on the main heatmap by clicking, with their molecular trajectories displayed in real-time line plots that enable direct comparison between cells experiencing different microenvironments. Selected cells can be individually edited, allowing users to manually adjust the concentration of any molecular species in a specific cell to test localized perturbations. The platform also provides figure export functionality, enabling users to save the current heatmap visualization in PNG and SVG formats for publication or further analysis. The comprehensive parameter control panel contains all eight primary rate constants with real-time adjustment capabilities, enabling immediate exploration of model sensitivity and behavior under different kinetic assumptions. This interactive capability transforms ECMSim into a digital laboratory environment where hypotheses can be tested and validated without the temporal and resource constraints of traditional experimental approaches.

\subsection*{\normalsize Spatiotemporal pattern formation and diffusion dynamics}

The platform's spatial modeling capabilities enable investigation of emergent tissue-level phenomena through the simulation of diffusion-driven pattern formation. Fig.~\ref{Figure 4} demonstrates three representative simulation scenarios illustrate the platform's ability to model spatially heterogeneous responses to localized stimuli. Case 01 represents the baseline condition with uniform zero input concentrations across all cells, resulting in spatially homogeneous concentrations dominated by stochastic fluctuations around the initial random values. The temporal evolution from iteration 0 to 500 shows gradual equilibration toward steady-state values with minimal spatial organization.

Case 02 demonstrates the response to localized TGF-$\beta$ stimulation applied in an X-shaped pattern through the brush selection interface. The TGF-$\beta$ input concentration of 1.0 in the selected region triggers robust intracellular signaling cascades that drive procollagen I (proCI) synthesis in the stimulated cells. During the early transient phase (e.g., iteration 100), the stimulated X region may display lower proCI concentration than the surrounding unstimulated cells because the intracellular signaling cascades are actively transitioning from the randomized initial state through receptor activation and kinase processing before ECM production increases; this lag reflects the time required for signals to propagate through the multi-step pathway hierarchy. At later time points (iterations 300--500), the stimulated region shows markedly higher concentrations as the signaling cascades have fully propagated to ECM production outputs. The temporal evolution reveals the gradual emergence of spatial patterns that reflect both the initial stimulation geometry and the subsequent diffusion of feedback molecules to neighboring unstimulated cells. By iteration 500, the spatial pattern exhibits clear concentration gradients extending from the initially stimulated regions. With continuous brush-applied stimulation, the system reaches a spatially heterogeneous steady state that maintains these gradients due to the persistent localized source; without sustained stimulation, the system would eventually approach a uniform steady state as diffusion equilibrates the concentration field, though this requires simulation times longer than the diffusion timescale across the full grid.

Case 03 illustrates the synergistic effects of combinatorial cytokine stimulation, where TGF-$\beta$, AngII, IL6, IL1, TNF-$\alpha$, and NE are simultaneously applied with input concentration of 1.0 in the selected X-shaped region. The multi-factor stimulation produces markedly enhanced proCI accumulation compared to TGF-$\beta$ alone, reflecting the cooperative interactions between different signaling pathways implemented in the network architecture. The spatial patterns exhibit both increased magnitude and extended spatial range, demonstrating how pathway crosstalk can amplify local stimuli into tissue-wide responses. The temporal progression shows accelerated kinetics and higher steady-state concentrations, illustrating the platform's capability to model complex combinatorial effects that are difficult to predict from individual pathway studies.

The diffusion dynamics across all three cases demonstrate the platform's ability to model realistic spatiotemporal behaviors, including signal propagation, boundary effects, and the emergence of concentration gradients. The periodic boundary conditions eliminate artificial edge effects while maintaining computational efficiency, enabling long-term simulations that capture both transient dynamics and steady-state behaviors. The visualization clearly shows how molecular diffusion creates smooth concentration gradients that extend beyond the initially stimulated regions, providing insights into the spatial scales of intercellular communication in cardiac tissue.

\subsection*{\normalsize Parameter sensitivity analysis}

To identify the dominant regulators of ECM remodeling outcomes, a one-at-a-time (OAT) sensitivity analysis was performed on the eight primary adjustable rate constants ($k_{\text{input}}$, $k_{\text{feedback}}$, $k_{\text{degradation}}$, $k_{\text{receptor}}$, $k_{\text{inhibition}}$, $k_{\text{activation}}$, $k_{\text{production}}$, and the integration time step $\Delta t$). Each parameter was perturbed by $\pm25\%$ and $\pm50\%$ from its default value while all other parameters were held constant. The sensitivity index was computed as the relative change in the transient area under the curve (AUC) of each ECM output during the dynamic accumulation phase ($T = 0$--$30$), where parameter differences manifest most clearly. Seven key ECM outputs were tracked: procollagen~I, procollagen~III, MMP-2, MMP-9, TIMP-1, $\alpha$-SMA, and fibronectin.

\begin{figure}[!ht]
	\includegraphics[width=1\textwidth]{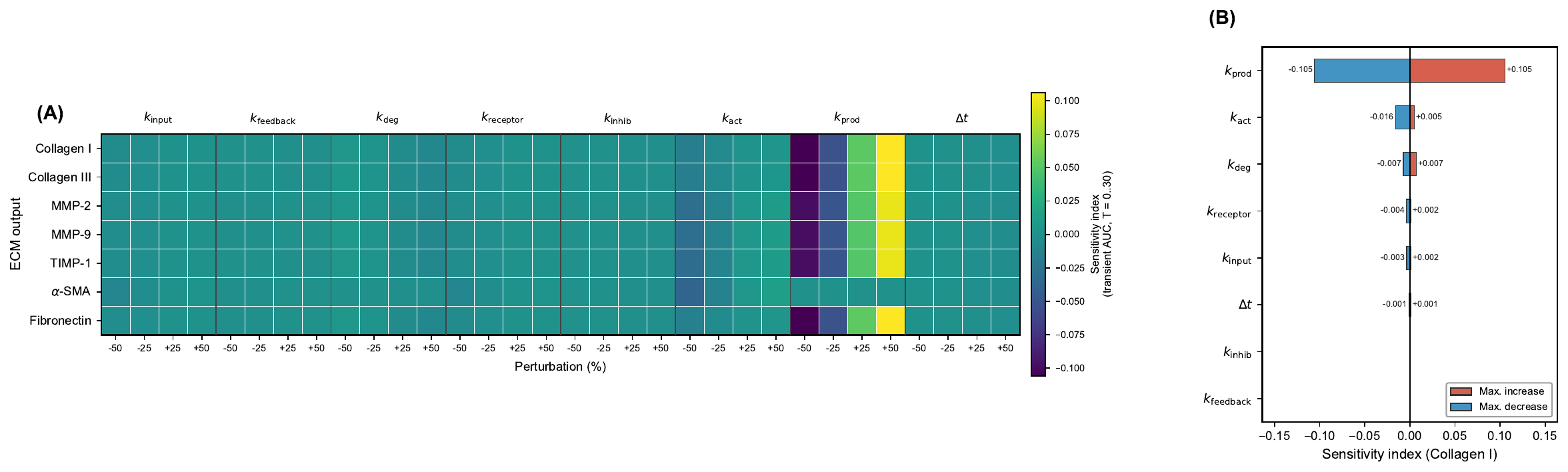}
	\caption{\footnotesize \textbf{Parameter sensitivity analysis of ECMSim.} (\textbf{A}) Heatmap of sensitivity indices for seven ECM outputs across eight model parameters perturbed by $\pm25\%$ and $\pm50\%$. Color intensity indicates the magnitude and direction of change in transient AUC relative to baseline. (\textbf{B}) Tornado chart showing the sensitivity range of each parameter for procollagen~I (Collagen~I), with bars indicating maximum increase (red) and maximum decrease (blue) across all perturbation levels.}
	\label{Figure 5}
\end{figure}

Fig.~\ref{Figure 5}A presents the sensitivity heatmap, where each column corresponds to a perturbation level of a given parameter and each row represents an ECM output. The analysis reveals that the ECM production rate ($k_{\text{production}}$) is by far the most influential parameter (mean $|SI| = 0.066$), followed by the activation rate ($k_{\text{activation}}$, mean $|SI| = 0.012$) and the degradation rate ($k_{\text{degradation}}$, mean $|SI| = 0.005$). The tornado chart (Fig.~\ref{Figure 5}B) confirms this hierarchy for procollagen~I, showing that $k_{\text{production}}$ perturbation of $\pm50\%$ produces sensitivity indices of approximately $\pm0.105$, while all other parameters produce indices below $\pm0.02$. The remaining parameters ($k_{\text{feedback}}$, $k_{\text{inhibition}}$, $k_{\text{input}}$, $k_{\text{receptor}}$, and $\Delta t$) have comparatively minor effects on ECM outputs, indicating that the model is robust to moderate variations in these parameters. These findings are consistent with the biological understanding that ECM accumulation in cardiac fibrosis is primarily governed by the rate of matrix protein production by activated fibroblasts and the balance between activation and degradation kinetics.

\captionsetup[figure]{labelformat=default}

\section*{\normalsize DISCUSSION}

ECMSim advances computational systems biology through its integration of high-performance numerical simulation with accessible web-based deployment, addressing fundamental limitations that have historically restricted the adoption of sophisticated modeling platforms in cardiovascular research. The platform's implementation of over 1 million coupled ordinary differential equations within a browser environment demonstrates the maturation of WebAssembly technology for scientific computing applications, achieving computational performance within 20\% of native implementations while eliminating the substantial barriers associated with specialized software installation and configuration. The mathematical framework underlying ECMSim advances beyond traditional approaches through its explicit modeling of spatial heterogeneity in cardiac fibroblast populations, incorporating both intracellular signaling complexity and intercellular communication through reaction-diffusion dynamics. This multi-scale integration addresses a critical gap identified in recent reviews of cardiac fibrosis modeling, {\color{blue}\cite{Zeigler2016,Zeigler2016review,Rogers2022,Rogers2022b,Rikard2019,Watts2023,Hays2025}} where spatial organization and cell-cell communication have been largely overlooked despite their central importance in pathological remodeling processes. The platform's brush-based spatial input system enables precise experimental design that mirrors clinical intervention strategies, such as localized drug delivery or regional mechanical stimulation protocols. \cite{Frangogiannis2019,Travers2016} The implementation of multiplicative regulatory interactions rather than simplified linear kinetics represents a significant advancement in capturing the complex nonlinear behaviors that characterize cellular signaling networks. This approach aligns with recent experimental findings demonstrating the prevalence of cooperative and competitive interactions in fibroblast signaling pathways, \cite{Tallquist2020,Fu2018} providing a more biologically realistic foundation for understanding emergent network behaviors and therapeutic intervention strategies.

The simulation results presented in Fig.~\ref{Figure 4} reveal fundamental principles of spatiotemporal pattern formation in cardiac tissue that have important implications for understanding pathological progression. The emergence of concentration gradients extending beyond initially stimulated regions demonstrates how localized inflammatory or mechanical stimuli can propagate across tissue domains through paracrine signaling mechanisms, consistent with recent experimental observations of fibrotic spread following myocardial infarction. \cite{Lighthouse2019,Frangogiannis2020} The platform's ability to capture both direct cellular responses and indirect paracrine effects addresses a fundamental challenge in translating molecular-scale experimental findings to tissue-level pathophysiology. Recent single-cell RNA sequencing studies have revealed substantial heterogeneity in cardiac fibroblast populations, \cite{Skelly2018,McLellan2020} and ECMSim's spatial framework provides a computational environment for investigating how this heterogeneity contributes to tissue-level function and dysfunction.

\subsection*{\normalsize Model applications and research capabilities}

ECMSim's comprehensive architecture enables systematic investigation of cardiac fibroblast behavior under diverse pathological conditions through its ability to simulate individual and combined stimuli across multiple spatial and temporal scales. The platform facilitates identification of key regulatory nodes and potential therapeutic targets through systematic perturbation studies where specific pathway components can be selectively activated or inhibited. The real-time parameter adjustment capabilities enable rapid exploration of parameter space to assess model robustness and identify critical rate constants that control system behavior. The spatial modeling capabilities provide unique insights into tissue-level phenomena that cannot be captured through traditional single-cell or homogeneous tissue approaches. The platform enables investigation of signaling wave propagation dynamics, boundary effects at tissue interfaces, and the emergence of spatial patterns in cellular responses that may contribute to pathological tissue architecture. The ability to apply spatially localized stimuli through the brush interface enables modeling of realistic pathological scenarios such as localized inflammatory foci, mechanical stress concentrations, or drug delivery patterns.

For research applications, ECMSim facilitates integration of experimental data from multiple sources and scales through its modular parameter structure and real-time calibration capabilities. Users can adjust model parameters to match their experimental observations, then use the calibrated model to predict cellular responses under conditions that have not been experimentally tested. The platform's ability to simulate combinatorial perturbations makes it particularly valuable for investigating drug interactions, optimizing multi-target therapeutic approaches, and understanding the complex interplay between different pathological stimuli. The educational applications of ECMSim extend beyond research to training environments where students and researchers can develop an intuitive understanding of complex signaling networks through direct manipulation and observation. The immediate visual feedback provided by the real-time visualizations enables users to develop mechanistic insights into how molecular interactions translate into tissue-level behaviors. The platform serves as a bridge between molecular-scale experimental observations and tissue-level pathological phenomena, providing a unified framework for understanding cardiac fibrosis across multiple scales of biological organization.

\subsection*{\normalsize Platform extensibility and community applications}

The modular architecture and open-source implementation of ECMSim enable broad applications beyond cardiac fibrosis through straightforward adaptation of the signaling network components and spatial parameters. The fundamental reaction-diffusion framework can accommodate diverse biological systems including cancer microenvironments, wound healing processes, developmental morphogenesis, and regenerative medicine applications. The web-based deployment eliminates installation barriers and enables collaborative model development across distributed research teams. The platform's extensibility is demonstrated through its capacity to incorporate new signaling pathways, molecular species, or regulatory interactions as biological knowledge advances. The object-oriented code structure facilitates community contributions and enables rapid integration of experimental findings into the computational framework. The real-time parameter adjustment capabilities support model validation against emerging experimental data and enable rapid hypothesis testing in response to new biological discoveries. ECMSim represents a significant advancement in accessible computational modeling of complex biological systems, demonstrating how modern web technologies can democratize sophisticated simulation capabilities previously available only through specialized software installations. The platform's combination of mathematical rigor, computational efficiency, and user accessibility creates new opportunities for interdisciplinary collaboration between computational and experimental researchers in cardiovascular biology and beyond.

\subsection*{\normalsize Therapeutic applications and drug discovery implications}

ECMSim's real-time parameter adjustment capabilities enable systematic investigation of therapeutic intervention strategies through virtual drug screening approaches. The platform's ability to simulate localized drug application through the spatial brush interface directly parallels emerging clinical approaches such as targeted nanoparticle delivery systems and localized gene therapy protocols \cite{Ishikawa2018}. The comprehensive pathway coverage enables investigation of both on-target and off-target effects, addressing a critical limitation in current drug development pipelines where systemic effects are often poorly predicted from single-pathway studies. \cite{Zhao2020} The platform's capacity to model combinatorial interventions provides particular value for developing multi-target therapeutic strategies that may be necessary for effective fibrosis treatment, suggesting that successful interventions will require simultaneous modulation of multiple pathways. ECMSim enables systematic exploration of combination therapy effects while accounting for potential antagonistic interactions that could limit therapeutic efficacy. The integration of mechanotransduction pathways enables investigation of mechanical intervention strategies, including cardiac rehabilitation protocols and device-based therapies. The platform can model how changes in mechanical loading conditions influence cellular signaling and matrix remodeling, providing insights into optimal timing and intensity of mechanical interventions. \cite{Coeyman2022}

\subsection*{\normalsize Technological advancement and accessibility democratization}

The WebAssembly-based architecture represents a significant technological advancement that addresses longstanding barriers to computational tool adoption in biological research. Traditional modeling platforms require specialized software installation, computational expertise, and often substantial computational resources that limit accessibility to researchers with advanced technical backgrounds. ECMSim's browser-based deployment eliminates these barriers while maintaining computational rigor, potentially democratizing access to sophisticated modeling capabilities across diverse research environments. The platform's modular architecture facilitates community-driven development through its object-oriented design and open-source implementation. This approach aligns with recent calls for increased reproducibility and transparency in computational biology, \cite{Stodden2018} enabling researchers to examine, modify, and extend the underlying mathematical models. The web-based deployment facilitates rapid dissemination of model updates and community contributions, addressing the version control challenges that often plague collaborative model development efforts. The real-time visualization capabilities provide immediate feedback that transforms the traditional modeling workflow from a batch-processing approach to an interactive exploration paradigm. This shift has important implications for educational applications, enabling students and trainees to develop intuitive understanding of complex signaling networks through direct manipulation and observation.

\subsection*{\normalsize Limitations and future development directions}

Several aspects of the current implementation warrant consideration for future enhancement. The ``tension'' variable in the current model represents mechanical stimulation that induces molecular- scale force across cell adhesion proteins and ion channels. Fibroblast mechanotransduction can driven by extracellular deformation and/or intracellular acto-myosin contractility, which activate downstream signaling cascades and ultimately influence gene expression through transcription factors such as YAP/TAZ and SRF. The ``tensionfb" feedback variable provides a way for our model to capture a sort of force propagation where tension from one cell pulling on their local matrix can act on other cells within small distances. Note we are not actually solving detailed traction force balances. A full finite element mechanical model coupled with the biochemical signaling network would be a valuable future extension that would enable modeling of physical deformations, fiber reorientation, and direct force transmission through the ECM. The current implementation supports periodic boundary conditions as the default, which effectively approximate an infinite tissue domain. Alternative boundary conditions, including fixed-value (Dirichlet) and no-flux (Neumann) conditions, would be valuable for modeling specific physiological scenarios such as tissue boundaries and infarct borders, and the modular code architecture readily supports their addition in future releases.

The two-dimensional spatial grid, while computationally efficient, represents a simplification of the complex three-dimensional architecture characteristic of cardiac tissue. Extension to three-dimensional geometries would enable investigation of transmural signaling gradients and the influence of tissue geometry on pattern formation, phenomena that are increasingly recognized as important factors in cardiac remodeling. \cite{Zeigler2016} The current implementation assumes homogeneous fibroblast populations, neglecting the phenotypic diversity revealed by recent single-cell studies. \cite{Skelly2018, Farbehi2019} Integration of cell-type-specific parameter sets based on single-cell transcriptomic data would enhance biological realism and enable investigation of how cellular heterogeneity contributes to tissue-level responses. This enhancement could incorporate stochastic elements to capture the noise inherent in cellular signaling systems, particularly relevant for low-abundance molecular species where fluctuations may significantly influence system behavior. The deterministic nature of the current mathematical framework may not adequately capture the stochastic fluctuations that characterize cellular signaling networks. Integration of stochastic differential equation approaches or agent-based modeling components could provide more realistic representations of cellular noise and its effects on pattern formation and signal propagation. \cite{Lim2015} Future developments could incorporate patient-specific parameterization through integration with clinical data sources, enabling personalized modeling approaches that account for individual genetic backgrounds, comorbidity profiles, and treatment histories. This capability would support precision medicine approaches by predicting individual responses to specific therapeutic interventions. \cite{Ashley2016}

\subsection*{\normalsize Broader implications and future applications}

The technological and methodological innovations demonstrated in ECMSim have implications extending beyond cardiac fibrosis to diverse areas of biomedical research. The reaction-diffusion framework underlying the platform is fundamentally applicable to any biological system involving spatial pattern formation and intercellular communication, including cancer progression, wound healing, developmental morphogenesis, tissue-based transcriptomics, \cite{Hays2023} and tissue engineering applications. \cite{Maini2012} The platform's success in democratizing access to sophisticated computational tools suggests potential for broader transformation of computational biology education and research practices. The combination of mathematical rigor with intuitive interfaces could accelerate the integration of quantitative approaches into traditionally experimental research programs, fostering interdisciplinary collaboration and accelerating scientific discovery. ECMSim's demonstration that complex biological systems can be effectively modeled within standard web browsers opens new possibilities for collaborative research across institutional boundaries. The platform could serve as a foundation for developing specialized modules targeting specific disease processes or therapeutic interventions, creating an ecosystem of interconnected modeling tools that collectively address the complexity of human disease. \cite{Hunter2013} The open-source architecture and community-driven development model position ECMSim to evolve continuously in response to advancing biological knowledge and technological capabilities. This approach ensures long-term sustainability while enabling rapid incorporation of emerging experimental findings into the computational framework, maintaining relevance as the field advances.

	\section*{\normalsize Resource Availability}
	\subsection*{\normalsize Lead contact}
	Further information and requests for resources should be directed to and will
	be fulfilled by the lead contact, William Richardson \texttt{\textcolor{blue}{wr013@uark.edu}} \label{leadcontact}.
	
	\subsection*{\normalsize Code availability}

	The original code is available at Richardson Lab in GitHub (\url{https://github.com/SysMechBioLab/ECMSim}).

	\section*{\normalsize Acknowledgments}
	This work was supported by the National Institutes of Health (NIGMS R01GM157589) and the Department of Defense (DEPSCoR FA9550-22-1-0379).
	
	\section*{\normalsize Author contributions}
	Model designing, methodology, coding, and writing--original draft: H.H.; Conceptualization, writing--review \& editing, funding acquisition, resources, and supervision: W.J.R.
	
	\section*{\normalsize Declaration of interests}
	No conflicts of interest to disclose.
	
	\vspace{1em}
	\noindent\rule{\linewidth}{0.4pt}
	\vspace{0.5em}
	
\begin{multicols}{2}[\section*{\normalsize REFERENCES}]
	\small 
	\renewcommand{\refname}{}
	\bibliographystyle{IEEEtran}  
	\bibliography{references}     
\end{multicols}

\end{document}